\documentclass[aps,prl,twocolumn,showpacs,groupedaddress,preprintnumbers]{revtex4-1}  
\usepackage{graphicx}  
\usepackage{dcolumn}   
\usepackage{bm}        
\usepackage{amssymb}   
\usepackage{braket}
\usepackage{amsmath}
\usepackage{color}  	
\usepackage[utf8]{inputenc}

\usepackage{breakurl} 

\newcommand{\bJ}{\mathbf{J}}
\newcommand{\bJt}{\mathbf{J}^{(0)}}
\newcommand{\bT}{\mathbf{T}}
\newcommand{\bI}{\mathbf{I}}
\newcommand{\bd}{\mathbf{d}}

\begin{document}

\preprint{MAN/HEP/2016/02 ~~ MCnet-16-04}
\title{Ordering multiple soft gluon emissions}
\author{Ren\'e \'Angeles Mart\'inez}
\author{Jeffrey R. Forshaw}
\author{Michael H. Seymour}
\affiliation{Consortium for Fundamental Physics,
  School of Physics and Astronomy,
  University of Manchester,
  Manchester M13 9PL,
  United Kingdom}
\date{\today}

\begin{abstract}
We present an expression for the QCD amplitude for a general hard scattering process with any number of soft gluon emissions, to one-loop accuracy. The amplitude is written in two different but equivalent ways: as a product of operators ordered in dipole transverse momentum and as a product of loop-expanded currents. We hope that these results will help in the development of an all-orders algorithm for multiple emissions that includes the full color structure and both the real and imaginary contributions to the amplitude.
\end{abstract}

\maketitle 

Soft gluon factorization (e.g. see \cite{Bassetto:1984ik}) is an important property of perturbative QCD. It is an essential ingredient in the construction of state-of-the-art Monte Carlo event generators \cite{Sjostrand:2014zea,Bahr:2008pv,*Bellm:2015jjp,Gleisberg:2008ta} and all-orders logarithmic resummations (e.g. see \cite{Luisoni:2015xha}), and it allows control over infrared poles in computations of cross sections at fixed order in the strong coupling (e.g. see \cite{Catani:1996vz}). In this Letter, we consider the one-loop amplitude for any number of soft gluon emissions off a general $n$-parton hard process. We work in the eikonal approximation for the coupling of the soft gluons to the hard partons and are able to derive an interesting identity in the limit where the gluons are ordered in softness. Specifically, we show that the $N$ emission amplitude may be written in two equivalent ways: either in terms of an ordered evolution of the hard scattering amplitude (in which intermediate infrared divergences cancel) or in terms of a product of loop-expanded emission operators acting on a loop-expanded matrix element (both of which are infrared divergent). It is our hope that Eq.~(\ref{eq:master1}), below, will form the basis for a future all-orders amplitude-level parton shower algorithm. This would be a major improvement over cross-section-level algorithms, as implemented in the existing parton shower event generators, not least because it would include full color evolution and Coulomb gluon exchange. Other work towards this goal can be found in 
Refs.~\cite{Platzer:2013fha,Sjodahl:2014opa,Gerwick:2014gya,Nagy:2015hwa}. As we will see, our calculations also indicate how successive real emissions constrain the intermediate (and finite) loop integrals by imposing an ordering fixed by the transverse momenta of adjacent real emissions, defined with respect to the directions of the partons involved in the virtual exchange.

In the ordered evolution approach, the one-loop amplitude for a total of $N$ soft gluon emissions, with four-momenta $q_i$, from a hard process with $n$ legs, with four-momenta $p_i$, is \cite{rene}
\begin{widetext}
\vspace*{-0mm}
\begin{eqnarray}
\label{eq:master1}
\ket{M^{(1)}_N} &=& 
\sum_{m=0}^{N} \sum_{i=2}^{p} \sum_{j=1}^{i-1} 
\, (g_s \mu^\epsilon)^{N-m} \, \bJt(q_N) \cdots \bJt(q_{m+1})\, \bI_{ij}(\tilde{q}_{m+1},\tilde{q}_{m}) \, \ket{M^{(0)}_m}  \\
&+&  \sum_{m=1}^N \sum_{j=1}^{n+m-1} \sum_{k=1}^{n+m-1} \, (g_s \mu^\epsilon)^{N-m} \,\bJt(q_N) \cdots \bJt(q_{m+1})
\, \bI_{n+m,j}(\tilde{q}_{m+1},q_{m}^{(jk)}) \, \bd_{jk}(q_{m}) \ket{M^{(0)}_{m-1}} ~,
\nonumber
\end{eqnarray}
\vspace*{-0mm}
\end{widetext}
\vspace*{-5.35mm}
where $p= n$ if $m=0$ or $p=n+m-1$ if $m \ge 1$. We assume $2 p_i \cdot p_j \sim Q^2$
for $i\ne j \le n$ and define $q_0 = Q$ and $q_{N+1} = 0$. 
We have defined the $m$-gluon amplitude
\begin{equation}
	\ket{M^{(0)}_m} = (g_s \mu^\epsilon)^{m} \,\bJt(q_m)  \cdots \bJt(q_1) \ket{M^{(0)}_0}~,
\end{equation}
where $\ket{M^{(0)}_0}$ is the hard scattering amplitude. The transverse momentum, defined with respect to partons $i$ and $j$, is
\begin{equation}
(q^{(ij)})^2 = 	\frac{2 \, q \cdot p_i \; q \cdot p_j}{p_i \cdot p_j}	,
\end{equation}
where $p_{n+a} = q_a$ and $\tilde{q} = q^{(ij)}$ when used in the argument of an $\bI_{ij}$ operator. The notation is such that all indices label partons such that parton $n+i$ indexes the soft gluon with momentum $q_i$. The one-loop insertion operator is
\begin{widetext}
\vspace*{-0mm}
\begin{equation}
\label{eq:insertop}
	\bI_{ij}(a,b) =  \frac{\alpha_s}{2 \pi}  \frac{c_{\Gamma}}{\epsilon^2}
 \, \bT_i \cdot \bT_j
\left[ \left( \frac{b^2}{4 \pi \mu^2} \right)^{-\epsilon}
\left(1 +i \pi \epsilon \,
	\tilde\delta_{ij}-\epsilon\ln\frac{2p_i\cdot p_j}{b^2}\right)
 - \left(\frac{a^2}{4\pi \mu^2} \right)^{-\epsilon} \, \left(1 +i \pi \epsilon \,
	\tilde\delta_{ij}-\epsilon\ln\frac{2p_i\cdot p_j}{a^2}\right) \right]~,
\end{equation}
\vspace*{-0mm}
\end{widetext}
\vspace*{-4.35mm}
where $\tilde{\delta}_{ij}=1$ if partons $i$ and $j$ are either both incoming or both outgoing and $\tilde\delta_{ij}=0$ otherwise and $c_{\Gamma} =  1 - \epsilon \gamma_E$, where $\gamma_E$ is the Euler-Mascheroni constant. This expression is accurate up to non-logarithmic terms of order of $\epsilon^0$ in the real part and the order of $\epsilon^1$ in the imaginary part. $\bI_{ij}(a,b)$ describes the evolution of partons $i$ and $j$ from transverse momentum $b$ to transverse momentum $a$ and, when both are non-zero, it is finite:
\begin{eqnarray}
	\bI_{ij}(a,b) &=&  \frac{\alpha_s}{2 \pi}
 \, \bT_i \cdot \bT_j
\biggl[ -\frac12\ln^2\frac{2p_i\cdot p_j}{b^2}
 +\frac12\ln^2\frac{2p_i\cdot p_j}{a^2}
\nonumber\\&&
 -i \pi \tilde\delta_{ij}\ln\frac{b^2}{a^2} \biggr]~,
\qquad (a^2,b^2>0)~.
\end{eqnarray}
The operator $\bd_{ij}(q)$ in Eq.~(\ref{eq:master1}),
\begin{equation}
\bd_{ij}(q) = \bT_j \left( \frac{p_j \cdot \varepsilon}{p_j \cdot q} -  \frac{p_i \cdot \varepsilon}{p_i \cdot q} \right)	,
\end{equation}
adds a soft gluon with four-momentum $q$ and polarization vector $\varepsilon$. The kinematic part (but not the color part) corresponds to an emission off the dipole formed by partons $i$ and $j$. Summing over dipoles is, by virtue of color conservation, equal to the soft gluon current, i.e.,
\begin{equation}
	\sum_{j} \bd_{ij}(q) = \sum_j \bT_j \frac{p_j \cdot \varepsilon}{p_j \cdot q} = \bJt(q) ~
\end{equation}
for any choice of parton $i$. The dependence upon parton $k$ in the argument of the insertion operator in the second line of Eq.~(\ref{eq:master1}) prevents the sum over $k$ converting the dipole operator $\bd_{jk}(q_m)$ into a soft gluon current.

Equation~(\ref{eq:master1}) assumes that the soft gluons couple to the hard partons in the eikonal approximation and that successive real emissions are ordered in softness, i.e., $q_{j+1} \sim \lambda q_j$ keeping the leading terms as $\lambda \to 0$. In particular, we do not make the eikonal approximation for emissions off prior soft emissions; i.e., we use the full triple gluon vertex for those. We have performed explicit Feynman diagram calculations, in the Feynman gauge, to confirm that Eq.~(\ref{eq:master1}) is correct in the case of one real emission and that it gives the correct imaginary part in the case of two real emissions. The case of two real emissions involves summing over many diagrams, and it is only after summing over diagrams that the amplitude is seen to be a simple function of the momenta $q_a^{(ij)}$ \cite{Angeles-Martinez:2015rna,rene}. The result to any number of loops seems likely to be a straightforward extension of Eq.~(\ref{eq:master1}); i.e., insert the one-loop insertion operator in all possible ways between a chain of $\bJt$ operators, taking care to treat specially the case where the exchange is between the previous soft emission and any other parton.

Equation~(\ref{eq:master1}) is written as a chain of operators with the virtual loop momentum bounded by the $q_a^{(ij)}$ of adjacent real emissions. It would be interesting to investigate further the connection between the present work and the dipole shower formalism  \cite{Gustafson:1987rq,Lonnblad:1992tz,Platzer:2009jq,Hoche:2015sya}, where the $q_a^{(ij)}$ are used to order emissions. As far as we can discern, the ordering prescribed by Eq.~(\ref{eq:master1}) is a new result. Note, in particular, how the second line (corresponding to a virtual exchange between the latest real emission and some other parton) is needed. This is because the transverse momentum that fixes the upper limit of the virtual loop integral would be zero if, as in the first line, it is evaluated with respect to the direction of the partons that exchange the virtual gluon. In this case, the relevant transverse momentum is instead defined by the direction of the  parton that  {\em emitted} the latest real emission. We stress that the emergence of this dipole-$k_T$ ordering is exact in the sense that it involves no approximations to the limits of the loop integrals. The only approximations are the eikonal approximation for  the vertices and propagators corresponding to gluon radiation off the original  hard partons and the assumption that successive real emissions are ordered in their softness.  

\begin{figure}[ht]
\centering
\includegraphics[width=0.29\textwidth]{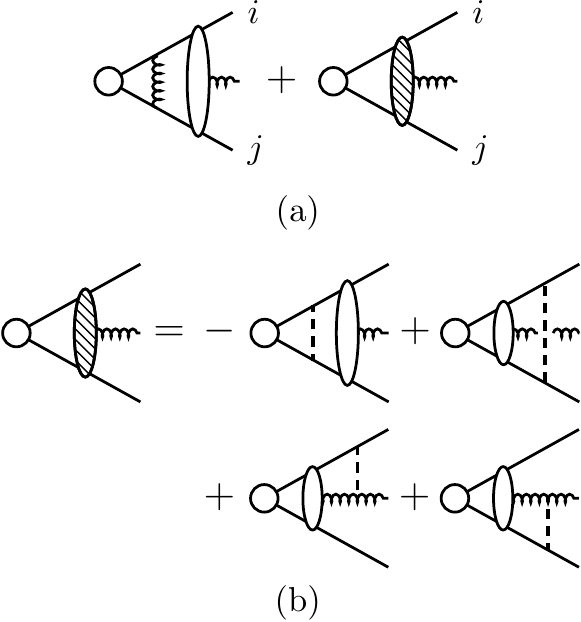}
\caption{(a) The one-loop, one-emission contributions in the loop-expansion approach. The unshaded oval indicates tree-level emission, while the shaded oval indicates emission at one loop [see Eq.~(\ref{eq:catgraz})]. (b) The key identity in demonstrating the equivalence of the loop-expansion approach and the evolution approach. The dotted lines indicate that the loop integral is to be cut off by the emitted gluon's transverse momentum, i.e.,  $\tilde{q}_m$ or $q_m^{(ij)}$. Note how the first term enters with a minus sign, so as to cancel the unwanted region of the first graph in (a).}
\label{fig:f1}
\end{figure}

The amplitude can also be written as a product of soft gluon emission operators acting on a dressed hard scattering amplitude:
\begin{equation} \label{newmaster}
\ket{M_N} = (g_s \mu^\epsilon)^{N} \, \bJ(q_N) \cdots \bJ(q_1)    \, \ket{M_0} ~, 
\end{equation}
where the soft gluon emission operator is now understood to have a loop expansion, i.e., at one-loop accuracy
$\bJ(q) = \bJ^{(0)}(q) + \bJ^{(1)}(q)$ with
\begin{equation}
	\bJ^{(1)}(q_{m+1}) = \frac{1}{2}\sum_{j=1}^{n+m} \sum_{k=1}^{n+m} \, 
        \bd_{jk}^{(1)}(q_{m+1}) \label{eq:oneloopJ}
\end{equation}
and
\vspace*{1mm}
\begin{eqnarray} \label{eq:catgraz}
\bd_{ij}^{(1)}(q_a) &=& 	\frac{\alpha_s}{2 \pi} \frac{c_\Gamma}{\epsilon^2}  \, \bT_{n+a} \cdot \bT_i \\ & \times &  \left( \frac{(q_a^{(ij)})^2 \; e^{-i \pi  \tilde\delta_{i(n+a)}} \, e^{-i \pi  \tilde\delta_{j (n+a)}}}{4\pi \mu^2 \, e^{-i \pi  \tilde\delta_{ij}}}\right)^{-\epsilon} \nonumber
\bd_{ij}(q_a)~. 
\end{eqnarray}
\vspace*{1mm}
This expression for the one-loop soft emission operator is equal to that previously derived by Catani and Grazzini \cite{Catani:2000pi}; see also \cite{Bern:1999ry,Feige:2014wja}.
Likewise, the hard scattering also has a loop expansion so that, at one loop,
$\ket{M_0} = \ket{M_0^{(0)}} + \ket{M_0^{(1)}}$, where
\begin{equation}
\ket{M_0^{(1)}} = \sum_{i=2}^{n} \sum_{j=1}^{i-1} \; \bI_{ij}(0,Q) \, \ket{M_0^{(0)}}~.
\end{equation}
In this way of organizing the perturbation series, the virtual corrections involving only hard partons are  factorized either into $\ket{M_0}$ or into the loop corrections to the emission operator. Because of the following result:
\begin{widetext}
\vspace*{-0mm}
\begin{eqnarray}	\label{eq:cool}
\bd_{ij}^{(1)}(q_{m})  \approx  
-\hspace{0.5mm} \bJt(q_{m}) \bI_{ij}(0,\tilde{q}_{m}) + \bI_{ij}(0,\tilde{q}_{m}) \bJt(q_{m}) 
+ \bI_{i,n+m}(0,q_{m}^{(ij)}) \bd_{ij}(q_{m})+\bI_{j,n+m}(0,q_{m}^{(ij)}) \bd_{ji}(q_{m})~, ~ ~~ 
\end{eqnarray}	
\vspace*{-0mm}
\end{widetext}
\vspace*{-0mm}
the one-loop part of Eq.~(\ref{newmaster}) is equal to Eq.~(\ref{eq:master1}). Intriguingly, Eq.~(\ref{eq:cool}), which makes the link between our amplitude-level evolution and the loop-expansion approach, would be exact in $\epsilon$ if we replaced the factor $1 +i \pi \epsilon \,\tilde\delta_{ij}$ in Eq.~(\ref{eq:insertop}) by $\cos(\pi \epsilon) + i \sin(\pi \epsilon \tilde\delta_{ij})$. Figure~\ref{fig:f1} illustrates Eq.~\eqref{eq:cool} graphically. Crucially, the equivalence between the two approaches hinges upon using the correct evolution variable. 
\vspace*{-0mm}


\begin{acknowledgements}
This work is supported by the Lancaster-Manchester-Sheffield
Consortium for Fundamental Physics STFC Grant No. ST/L000520/1 and by the Mexican National Council
of Science and Technology CONACyT Grant  No. 310259.
\end{acknowledgements}

\bibliography{refs}

\begin{thebibliography}{20}%
\makeatletter
\providecommand \@ifxundefined [1]{%
 \@ifx{#1\undefined}
}%
\providecommand \@ifnum [1]{%
 \ifnum #1\expandafter \@firstoftwo
 \else \expandafter \@secondoftwo
 \fi
}%
\providecommand \@ifx [1]{%
 \ifx #1\expandafter \@firstoftwo
 \else \expandafter \@secondoftwo
 \fi
}%
\providecommand \natexlab [1]{#1}%
\providecommand \enquote  [1]{``#1''}%
\providecommand \bibnamefont  [1]{#1}%
\providecommand \bibfnamefont [1]{#1}%
\providecommand \citenamefont [1]{#1}%
\providecommand \href@noop [0]{\@secondoftwo}%
\providecommand \href [0]{\begingroup \@sanitize@url \@href}%
\providecommand \@href[1]{\@@startlink{#1}\@@href}%
\providecommand \@@href[1]{\endgroup#1\@@endlink}%
\providecommand \@sanitize@url [0]{\catcode `\\12\catcode `\$12\catcode
  `\&12\catcode `\#12\catcode `\^12\catcode `\_12\catcode `\%12\relax}%
\providecommand \@@startlink[1]{}%
\providecommand \@@endlink[0]{}%
\providecommand \url  [0]{\begingroup\@sanitize@url \@url }%
\providecommand \@url [1]{\endgroup\@href {#1}{\urlprefix }}%
\providecommand \urlprefix  [0]{URL }%
\providecommand \Eprint [0]{\href }%
\providecommand \doibase [0]{http://dx.doi.org/}%
\providecommand \selectlanguage [0]{\@gobble}%
\providecommand \bibinfo  [0]{\@secondoftwo}%
\providecommand \bibfield  [0]{\@secondoftwo}%
\providecommand \translation [1]{[#1]}%
\providecommand \BibitemOpen [0]{}%
\providecommand \bibitemStop [0]{}%
\providecommand \bibitemNoStop [0]{.\EOS\space}%
\providecommand \EOS [0]{\spacefactor3000\relax}%
\providecommand \BibitemShut  [1]{\csname bibitem#1\endcsname}%
\let\auto@bib@innerbib\@empty
\bibitem [{\citenamefont {Bassetto}\ \emph {et~al.}(1983)\citenamefont
  {Bassetto}, \citenamefont {Ciafaloni},\ and\ \citenamefont
  {Marchesini}}]{Bassetto:1984ik}%
  \BibitemOpen
  \bibfield  {author} {\bibinfo {author} {\bibfnamefont {A.}~\bibnamefont
  {Bassetto}}, \bibinfo {author} {\bibfnamefont {M.}~\bibnamefont {Ciafaloni}},
  \ and\ \bibinfo {author} {\bibfnamefont {G.}~\bibnamefont {Marchesini}},\
  }\href {\doibase 10.1016/0370-1573(83)90083-2} {\bibfield  {journal}
  {\bibinfo  {journal} {Phys. Rept.}\ }\textbf {\bibinfo {volume} {100}},\
  \bibinfo {pages} {201} (\bibinfo {year} {1983})}\BibitemShut {NoStop}%
\bibitem [{\citenamefont {Sj{\"o}strand}\ \emph {et~al.}(2015)\citenamefont
  {Sj{\"o}strand}, \citenamefont {Ask}, \citenamefont {Christiansen},
  \citenamefont {Corke}, \citenamefont {Desai}, \citenamefont {Ilten},
  \citenamefont {Mrenna}, \citenamefont {Prestel}, \citenamefont {Rasmussen},\
  and\ \citenamefont {Skands}}]{Sjostrand:2014zea}%
  \BibitemOpen
  \bibfield  {author} {\bibinfo {author} {\bibfnamefont {T.}~\bibnamefont
  {Sj{\"o}strand}}, \bibinfo {author} {\bibfnamefont {S.}~\bibnamefont {Ask}},
  \bibinfo {author} {\bibfnamefont {J.~R.}\ \bibnamefont {Christiansen}},
  \bibinfo {author} {\bibfnamefont {R.}~\bibnamefont {Corke}}, \bibinfo
  {author} {\bibfnamefont {N.}~\bibnamefont {Desai}}, \bibinfo {author}
  {\bibfnamefont {P.}~\bibnamefont {Ilten}}, \bibinfo {author} {\bibfnamefont
  {S.}~\bibnamefont {Mrenna}}, \bibinfo {author} {\bibfnamefont
  {S.}~\bibnamefont {Prestel}}, \bibinfo {author} {\bibfnamefont {C.~O.}\
  \bibnamefont {Rasmussen}}, \ and\ \bibinfo {author} {\bibfnamefont {P.~Z.}\
  \bibnamefont {Skands}},\ }\href {\doibase 10.1016/j.cpc.2015.01.024}
  {\bibfield  {journal} {\bibinfo  {journal} {Comput. Phys. Commun.}\ }\textbf
  {\bibinfo {volume} {191}},\ \bibinfo {pages} {159} (\bibinfo {year}
  {2015})},\ \Eprint {http://arxiv.org/abs/1410.3012} {arXiv:1410.3012
  [hep-ph]} \BibitemShut {NoStop}%
\bibitem [{\citenamefont {Bahr}\ \emph {et~al.}(2008)\citenamefont {Bahr} \emph
  {et~al.}}]{Bahr:2008pv}%
  \BibitemOpen
  \bibfield  {author} {\bibinfo {author} {\bibfnamefont {M.}~\bibnamefont
  {Bahr}} \emph {et~al.},\ }\href {\doibase 10.1140/epjc/s10052-008-0798-9}
  {\bibfield  {journal} {\bibinfo  {journal} {Eur. Phys. J.}\ }\textbf
  {\bibinfo {volume} {C58}},\ \bibinfo {pages} {639} (\bibinfo {year}
  {2008})},\ \Eprint {http://arxiv.org/abs/0803.0883} {arXiv:0803.0883
  [hep-ph]} \BibitemShut {NoStop}%
\bibitem [{\citenamefont {Bellm}\ \emph {et~al.}(2015)\citenamefont {Bellm}
  \emph {et~al.}}]{Bellm:2015jjp}%
  \BibitemOpen
  \bibfield  {author} {\bibinfo {author} {\bibfnamefont {J.}~\bibnamefont
  {Bellm}} \emph {et~al.},\ }\href@noop {} {\  (\bibinfo {year} {2015})},\
  \Eprint {http://arxiv.org/abs/1512.01178} {arXiv:1512.01178 [hep-ph]}
  \BibitemShut {NoStop}%
\bibitem [{\citenamefont {Gleisberg}\ \emph {et~al.}(2009)\citenamefont
  {Gleisberg}, \citenamefont {H{\"o}che}, \citenamefont {Krauss}, \citenamefont
  {Sch{\"o}nherr}, \citenamefont {Schumann}, \citenamefont {Siegert},\ and\
  \citenamefont {Winter}}]{Gleisberg:2008ta}%
  \BibitemOpen
  \bibfield  {author} {\bibinfo {author} {\bibfnamefont {T.}~\bibnamefont
  {Gleisberg}}, \bibinfo {author} {\bibfnamefont {S.}~\bibnamefont
  {H{\"o}che}}, \bibinfo {author} {\bibfnamefont {F.}~\bibnamefont {Krauss}},
  \bibinfo {author} {\bibfnamefont {M.}~\bibnamefont {Sch{\"o}nherr}}, \bibinfo
  {author} {\bibfnamefont {S.}~\bibnamefont {Schumann}}, \bibinfo {author}
  {\bibfnamefont {F.}~\bibnamefont {Siegert}}, \ and\ \bibinfo {author}
  {\bibfnamefont {J.}~\bibnamefont {Winter}},\ }\href {\doibase
  10.1088/1126-6708/2009/02/007} {\bibfield  {journal} {\bibinfo  {journal}
  {JHEP}\ }\textbf {\bibinfo {volume} {02}},\ \bibinfo {pages} {007} (\bibinfo
  {year} {2009})},\ \Eprint {http://arxiv.org/abs/0811.4622} {arXiv:0811.4622
  [hep-ph]} \BibitemShut {NoStop}%
\bibitem [{\citenamefont {Luisoni}\ and\ \citenamefont
  {Marzani}(2015)}]{Luisoni:2015xha}%
  \BibitemOpen
  \bibfield  {author} {\bibinfo {author} {\bibfnamefont {G.}~\bibnamefont
  {Luisoni}}\ and\ \bibinfo {author} {\bibfnamefont {S.}~\bibnamefont
  {Marzani}},\ }\href {\doibase 10.1088/0954-3899/42/10/103101} {\bibfield
  {journal} {\bibinfo  {journal} {J. Phys.}\ }\textbf {\bibinfo {volume}
  {G42}},\ \bibinfo {pages} {103101} (\bibinfo {year} {2015})},\ \Eprint
  {http://arxiv.org/abs/1505.04084} {arXiv:1505.04084 [hep-ph]} \BibitemShut
  {NoStop}%
\bibitem [{\citenamefont {Catani}\ and\ \citenamefont
  {Seymour}(1997)}]{Catani:1996vz}%
  \BibitemOpen
  \bibfield  {author} {\bibinfo {author} {\bibfnamefont {S.}~\bibnamefont
  {Catani}}\ and\ \bibinfo {author} {\bibfnamefont {M.~H.}\ \bibnamefont
  {Seymour}},\ }\href {\doibase 10.1016/S0550-3213(96)00589-5} {\bibfield
  {journal} {\bibinfo  {journal} {Nucl. Phys.}\ }\textbf {\bibinfo {volume}
  {B485}},\ \bibinfo {pages} {291} (\bibinfo {year} {1997})},\ \bibinfo {note}
  {[Erratum: Nucl. Phys.B510,503(1998)]},\ \Eprint
  {http://arxiv.org/abs/hep-ph/9605323} {arXiv:hep-ph/9605323 [hep-ph]}
  \BibitemShut {NoStop}%
\bibitem [{\citenamefont {Pl{\"a}tzer}(2014)}]{Platzer:2013fha}%
  \BibitemOpen
  \bibfield  {author} {\bibinfo {author} {\bibfnamefont {S.}~\bibnamefont
  {Pl{\"a}tzer}},\ }\href {\doibase 10.1140/epjc/s10052-014-2907-2} {\bibfield
  {journal} {\bibinfo  {journal} {Eur. Phys. J.}\ }\textbf {\bibinfo {volume}
  {C74}},\ \bibinfo {pages} {2907} (\bibinfo {year} {2014})},\ \Eprint
  {http://arxiv.org/abs/1312.2448} {arXiv:1312.2448 [hep-ph]} \BibitemShut
  {NoStop}%
\bibitem [{\citenamefont {Sj{\"o}dahl}(2015)}]{Sjodahl:2014opa}%
  \BibitemOpen
  \bibfield  {author} {\bibinfo {author} {\bibfnamefont {M.}~\bibnamefont
  {Sj{\"o}dahl}},\ }\href {\doibase 10.1140/epjc/s10052-015-3417-6} {\bibfield
  {journal} {\bibinfo  {journal} {Eur. Phys. J.}\ }\textbf {\bibinfo {volume}
  {C75}},\ \bibinfo {pages} {236} (\bibinfo {year} {2015})},\ \Eprint
  {http://arxiv.org/abs/1412.3967} {arXiv:1412.3967 [hep-ph]} \BibitemShut
  {NoStop}%
\bibitem [{\citenamefont {Gerwick}\ \emph {et~al.}(2015)\citenamefont
  {Gerwick}, \citenamefont {H{\"o}che}, \citenamefont {Marzani},\ and\
  \citenamefont {Schumann}}]{Gerwick:2014gya}%
  \BibitemOpen
  \bibfield  {author} {\bibinfo {author} {\bibfnamefont {E.}~\bibnamefont
  {Gerwick}}, \bibinfo {author} {\bibfnamefont {S.}~\bibnamefont {H{\"o}che}},
  \bibinfo {author} {\bibfnamefont {S.}~\bibnamefont {Marzani}}, \ and\
  \bibinfo {author} {\bibfnamefont {S.}~\bibnamefont {Schumann}},\ }\href
  {\doibase 10.1007/JHEP02(2015)106} {\bibfield  {journal} {\bibinfo  {journal}
  {JHEP}\ }\textbf {\bibinfo {volume} {02}},\ \bibinfo {pages} {106} (\bibinfo
  {year} {2015})},\ \Eprint {http://arxiv.org/abs/1411.7325} {arXiv:1411.7325
  [hep-ph]} \BibitemShut {NoStop}%
\bibitem [{\citenamefont {Nagy}\ and\ \citenamefont
  {Soper}(2015)}]{Nagy:2015hwa}%
  \BibitemOpen
  \bibfield  {author} {\bibinfo {author} {\bibfnamefont {Z.}~\bibnamefont
  {Nagy}}\ and\ \bibinfo {author} {\bibfnamefont {D.~E.}\ \bibnamefont
  {Soper}},\ }\href {\doibase 10.1007/JHEP07(2015)119} {\bibfield  {journal}
  {\bibinfo  {journal} {JHEP}\ }\textbf {\bibinfo {volume} {07}},\ \bibinfo
  {pages} {119} (\bibinfo {year} {2015})},\ \Eprint
  {http://arxiv.org/abs/1501.00778} {arXiv:1501.00778 [hep-ph]} \BibitemShut
  {NoStop}%
\bibitem [{\citenamefont {{{\'A}ngeles Mart{\'i}nez}}(2015)}]{rene}%
  \BibitemOpen
  \bibfield  {author} {\bibinfo {author} {\bibfnamefont {R.}~\bibnamefont
  {{{\'A}ngeles Mart{\'i}nez}}},\ }\emph {\bibinfo {title} {Coulomb gluons and
  the ordering variable}},\ \href@noop {} {Ph.D. thesis},\ \bibinfo {address}
  {The University of Manchester} (\bibinfo {year} {2015})\BibitemShut {NoStop}%
\bibitem [{\citenamefont {{\'A}ngeles~Mart{\'i}nez}\ \emph
  {et~al.}(2015)\citenamefont {{\'A}ngeles~Mart{\'i}nez}, \citenamefont
  {Forshaw},\ and\ \citenamefont {Seymour}}]{Angeles-Martinez:2015rna}%
  \BibitemOpen
  \bibfield  {author} {\bibinfo {author} {\bibfnamefont {R.}~\bibnamefont
  {{\'A}ngeles~Mart{\'i}nez}}, \bibinfo {author} {\bibfnamefont {J.~R.}\
  \bibnamefont {Forshaw}}, \ and\ \bibinfo {author} {\bibfnamefont {M.~H.}\
  \bibnamefont {Seymour}},\ }\href {\doibase 10.1007/JHEP12(2015)091}
  {\bibfield  {journal} {\bibinfo  {journal} {JHEP}\ }\textbf {\bibinfo
  {volume} {12}},\ \bibinfo {pages} {091} (\bibinfo {year} {2015})},\ \Eprint
  {http://arxiv.org/abs/1510.07998} {arXiv:1510.07998 [hep-ph]} \BibitemShut
  {NoStop}%
\bibitem [{\citenamefont {Gustafson}\ and\ \citenamefont
  {Pettersson}(1988)}]{Gustafson:1987rq}%
  \BibitemOpen
  \bibfield  {author} {\bibinfo {author} {\bibfnamefont {G.}~\bibnamefont
  {Gustafson}}\ and\ \bibinfo {author} {\bibfnamefont {U.}~\bibnamefont
  {Pettersson}},\ }\href {\doibase 10.1016/0550-3213(88)90441-5} {\bibfield
  {journal} {\bibinfo  {journal} {Nucl. Phys.}\ }\textbf {\bibinfo {volume}
  {B306}},\ \bibinfo {pages} {746} (\bibinfo {year} {1988})}\BibitemShut
  {NoStop}%
\bibitem [{\citenamefont {L{\"o}nnblad}(1992)}]{Lonnblad:1992tz}%
  \BibitemOpen
  \bibfield  {author} {\bibinfo {author} {\bibfnamefont {L.}~\bibnamefont
  {L{\"o}nnblad}},\ }\href {\doibase 10.1016/0010-4655(92)90068-A} {\bibfield
  {journal} {\bibinfo  {journal} {Comput. Phys. Commun.}\ }\textbf {\bibinfo
  {volume} {71}},\ \bibinfo {pages} {15} (\bibinfo {year} {1992})}\BibitemShut
  {NoStop}%
\bibitem [{\citenamefont {Pl{\"a}tzer}\ and\ \citenamefont
  {Gieseke}(2011)}]{Platzer:2009jq}%
  \BibitemOpen
  \bibfield  {author} {\bibinfo {author} {\bibfnamefont {S.}~\bibnamefont
  {Pl{\"a}tzer}}\ and\ \bibinfo {author} {\bibfnamefont {S.}~\bibnamefont
  {Gieseke}},\ }\href {\doibase 10.1007/JHEP01(2011)024} {\bibfield  {journal}
  {\bibinfo  {journal} {JHEP}\ }\textbf {\bibinfo {volume} {01}},\ \bibinfo
  {pages} {024} (\bibinfo {year} {2011})},\ \Eprint
  {http://arxiv.org/abs/0909.5593} {arXiv:0909.5593 [hep-ph]} \BibitemShut
  {NoStop}%
\bibitem [{\citenamefont {H{\"o}che}\ and\ \citenamefont
  {Prestel}(2015)}]{Hoche:2015sya}%
  \BibitemOpen
  \bibfield  {author} {\bibinfo {author} {\bibfnamefont {S.}~\bibnamefont
  {H{\"o}che}}\ and\ \bibinfo {author} {\bibfnamefont {S.}~\bibnamefont
  {Prestel}},\ }\href {\doibase 10.1140/epjc/s10052-015-3684-2} {\bibfield
  {journal} {\bibinfo  {journal} {Eur. Phys. J.}\ }\textbf {\bibinfo {volume}
  {C75}},\ \bibinfo {pages} {461} (\bibinfo {year} {2015})},\ \Eprint
  {http://arxiv.org/abs/1506.05057} {arXiv:1506.05057 [hep-ph]} \BibitemShut
  {NoStop}%
\bibitem [{\citenamefont {Catani}\ and\ \citenamefont
  {Grazzini}(2000)}]{Catani:2000pi}%
  \BibitemOpen
  \bibfield  {author} {\bibinfo {author} {\bibfnamefont {S.}~\bibnamefont
  {Catani}}\ and\ \bibinfo {author} {\bibfnamefont {M.}~\bibnamefont
  {Grazzini}},\ }\href {\doibase 10.1016/S0550-3213(00)00572-1} {\bibfield
  {journal} {\bibinfo  {journal} {Nucl. Phys.}\ }\textbf {\bibinfo {volume}
  {B591}},\ \bibinfo {pages} {435} (\bibinfo {year} {2000})},\ \Eprint
  {http://arxiv.org/abs/hep-ph/0007142} {arXiv:hep-ph/0007142 [hep-ph]}
  \BibitemShut {NoStop}%
\bibitem [{\citenamefont {Bern}\ \emph {et~al.}(1999)\citenamefont {Bern},
  \citenamefont {Del~Duca}, \citenamefont {Kilgore},\ and\ \citenamefont
  {Schmidt}}]{Bern:1999ry}%
  \BibitemOpen
  \bibfield  {author} {\bibinfo {author} {\bibfnamefont {Z.}~\bibnamefont
  {Bern}}, \bibinfo {author} {\bibfnamefont {V.}~\bibnamefont {Del~Duca}},
  \bibinfo {author} {\bibfnamefont {W.~B.}\ \bibnamefont {Kilgore}}, \ and\
  \bibinfo {author} {\bibfnamefont {C.~R.}\ \bibnamefont {Schmidt}},\ }\href
  {\doibase 10.1103/PhysRevD.60.116001} {\bibfield  {journal} {\bibinfo
  {journal} {Phys. Rev.}\ }\textbf {\bibinfo {volume} {D60}},\ \bibinfo {pages}
  {116001} (\bibinfo {year} {1999})},\ \Eprint
  {http://arxiv.org/abs/hep-ph/9903516} {arXiv:hep-ph/9903516 [hep-ph]}
  \BibitemShut {NoStop}%
\bibitem [{\citenamefont {Feige}\ and\ \citenamefont
  {Schwartz}(2014)}]{Feige:2014wja}%
  \BibitemOpen
  \bibfield  {author} {\bibinfo {author} {\bibfnamefont {I.}~\bibnamefont
  {Feige}}\ and\ \bibinfo {author} {\bibfnamefont {M.~D.}\ \bibnamefont
  {Schwartz}},\ }\href {\doibase 10.1103/PhysRevD.90.105020} {\bibfield
  {journal} {\bibinfo  {journal} {Phys. Rev.}\ }\textbf {\bibinfo {volume}
  {D90}},\ \bibinfo {pages} {105020} (\bibinfo {year} {2014})},\ \Eprint
  {http://arxiv.org/abs/1403.6472} {arXiv:1403.6472 [hep-ph]} \BibitemShut
  {NoStop}%
\end{thebibliography}%

\end{document}